\documentclass{camera}
\usepackage{graphicx}  

\begin{document}

%
\title{BIMODALITY AND LATENT HEAT OF GOLD NUCLEI}

%
\author{E. Bonnet\\for the INDRA and ALADIN Collaborations}

%
\organization{GANIL, (DSM-CEA/CNRS/IN2P3), 
F-14076 Caen cedex, France}
\maketitle

\begin{abstract}
Peripheral collisions give access to a set of events where hot quasi-projectile (QP) sources
are produced exploring a large range of excitation energy. In this range evaporation and multifragmentation
are both observed for a similar charge/size of the sources. In this work, this experimental fact will be described in terms of 
coexistence of two classes of events in the first order phase transition formalism. 
For this, a detailed study of the experimental correlation between the size/charge of the biggest cluster/fragment ($Z_{1}$) and the excitation energy ($E^{*}$) is made. 
Making the parallel with the same correlation derived from the canonical ensemble description, a first value of the latent heat and boundaries of spinodal and coexistence 
zone are extracted.
\end{abstract}

%

\section{Introduction}

The distribution of an order parameter of a finite system which passes through a first order phase transition is expected to
be bimodal~\cite{Cho04}. For hot nuclei, this bimodality is an echo of two classes of events which undergo evaporation or multifragmentation. 
To be observed the populations of the two de-excitation channels has to be roughly equivalent. This is conditioned first of all by a sufficient exploration of the phase diagram.
Concerning this point, semi-peripheral collisions of heavy ion collisions produce in the outgoing channel hot quasi-projectile (QP) nuclei with a large range of excitation energy. 
The second condition to observe bimodality is the equivalent population of the two phases. In heavy ion collisions, impact parameter distributions mainly govern the distribution
of dissipated energy and largely favour low dissipation - big residue - evaporation channel class of events. In order to get rid of this experimental fact, previous studies~\cite{Pichon} 
proposed to sample QP events with the transverse energy of the light charged particle emitted on the quasi-target side. This dissipation sampling was related to canonical
temperature sampling one can perform in canonical ensemble to define the transition region where bimodality is observed. By this way, they indeed observe a clear transition between 
a dominant evaporation-like decay mode, with the biggest cluster much heavier than the second one, and a dominant fragmentation mode, with the two heaviest 
fragments of similar size. A similar behavior has been reported in~\cite{Bruno}.  Different physical scenarios have been invoked to interpret the phenomenon: finite-system counterpart of the nuclear 
matter liquid-gas phase transition~\cite{Pichon,Renorm}, Jacobi transition of highly deformed systems~\cite{Lopez}, self-organized criticality induced by nucleon-nucleon 
collisions~\cite{Lefevre}. In~\cite{Bruno}, the two decay modes were associated to different excitation energies, suggesting a temperature-induced transition 
with non-zero latent heat. The qualitative agreement between refs.~\cite{Pichon,Bruno} suggests that bimodality is a generic phenomenon. 
However, differences between the two data sets subsist, and trigger or selection bias cannot be excluded. To disentangle between the different scenarios, 
it is necessary to control the role of the entrance channel dynamics and establish if the transition is of thermal character.

For Au+Au collisions between 60 and 100 A.MeV performed at GSI and collected with the INDRA multidetector, a set of events covering an excitation energy range of [1;10] A.MeV
with size/charge variation of the reconstructed source around 10\% can be selected. In refs~\cite{Pichon,BonnetNPA}, two methods of selection based on the kinematical properties of events 
are proposed in order to remove from the studied set of events, so called dynamical events containing a contribution from the mid-rapidity region. In~\cite{BonnetPRL}, a detailed
study of the experimental correlation between the charge/size of the biggest fragment/cluster ($Z_{1}$) and the excitation energy ($E^{*}$) is performed for these sets of events.

\section{Bimodality and latent heat of Gold nuclei}

\begin{figure}[!h]
\begin{center}
\includegraphics[scale=0.70]{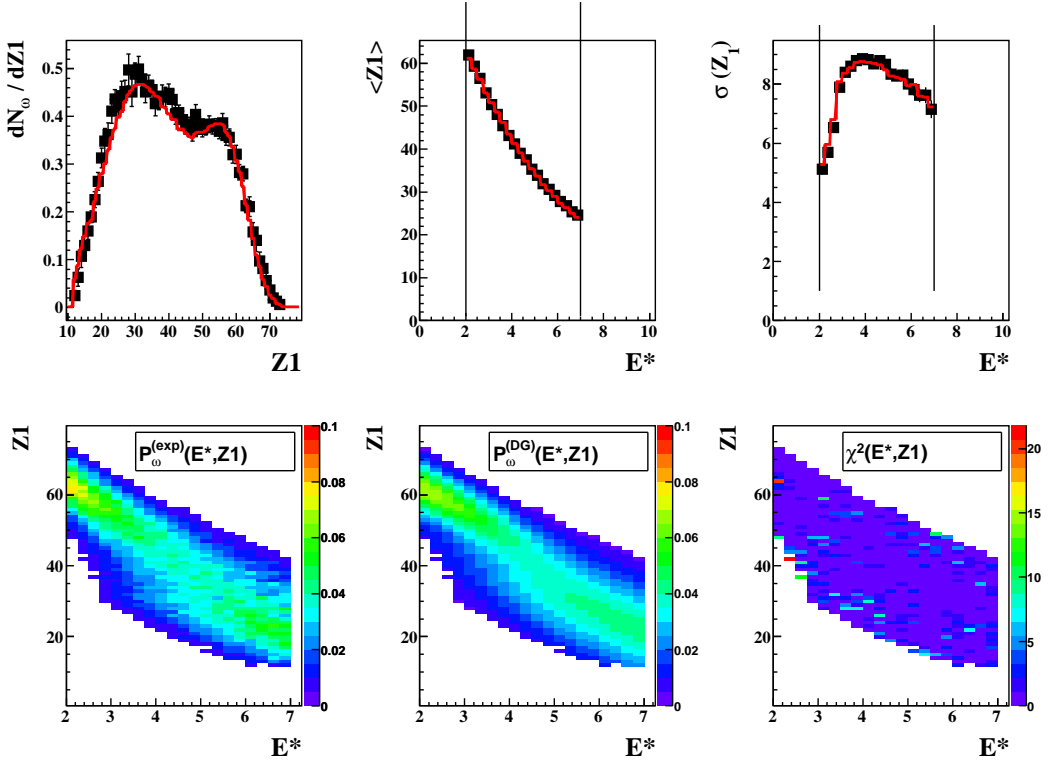}
\caption{\it top panel, from left to right: renormalized distribution of the charge of the biggest fragment ($Z_{1}$), evolution of the mean value and the standard
deviation of $Z_{1}$ with the excitation energy ($E^{*}$); the black squares correspond to the data, the red line to the canonical ensemble
function; bottom panel, from left to right: renormalized 2D experimental and canonical correlation between $Z_{1}$ and $E^{*}$, Chi2 estimator obtained from the fit procedure for each cell of the correlation.}
\label{fig_01}
\end{center}
\end{figure}

The canonical correlation of two order paremeters can be defined as $P_{can}(E^{*},Z_{1}) = W(E^{*},Z_{1})\;e^{-\beta E^{*}}\mathcal{Z}_{\beta}^{-1}$ where W is the density of states,
$\beta$ is the canonical temperature and $\mathcal{Z}_{\beta}$ is the associated partition sum. The bimodal shape of this correlation is directly connected to the
residual convex intruder in the entropy $d^{2}S/d E^{2}>0$ with $S(E^{*},Z_{1})=ln\;W(E^{*},Z_{1})$~\cite{Cho04}.
Studying the shape of $P_{can}$, one can localize two generic regions on the phase diagram: the coexistence region correponding to the region between its two maxima (and directly the
heat capacity which is simply their relative distance on the $E^*$ direction), and the spinodal region
where the second derivative of $P_{can}$ is positive ($d^{2}S/d E^{2} = d^{2} ln\;P_{can}/d E^{2}$).
This theoretical description has to be brought closer to experimental case and to access information from the bimodality signal, a procedure has been proposed in
ref.~\cite{Renorm}. In experimental data, the distribution of excitation energy is mainly governed by the impact parameter distribution, low dissipations are favoured. Measured
excitation energy distributions cannot be directly related to what we can expect for the distribution of an order parameter. The idea is to study simultaneously two observables
$E^{*}$ and $Z_{1}$ which are good candidates for order parameters of the phase transition which occurs in nuclei~\cite{Zmax}. Considering that the density of states of a nuclear 
system exploring a large range of excitation energy exhibits the generic convex intruder, we can expect to underline consequent bimodality by getting rid of entrance channel
effects. We consider the experimental correlation $P_{exp}(E^{*},Z_{1}) \propto W(E^{*},Z_{1})g_{exp}(E^{*})$ the simple product between the density of states and
experimental weight which includes entrance channel effects, detection and selection bias. To put face to face $P_{exp}$ and $P_{can}$, a renormalization of the
two correlations under the constraint of a flat distribution of excitation energy is made. It gives $P^{\omega}_{can,exp}(E^{*},Z_{1}) = P_{can,exp}(E^{*},Z_{1})/P_{can,exp}(E*)$.
Boltzmann factor and partition sum for the canonical expression and $g_{exp}$ for the experimental expression do not contribute anymore. We can focus directly on the density of
states: $W(E^{*},Z_{1})/W(E^{*})$, and compare the two normalized correlations. The major assumption to apply this procedure is the $E^*$ dependence only of the experimental
weight $g_{exp}$. This means that for a given value of $E^*$ the associated distribution of $Z_{1}$ is not biased by the entrance channel and/or detection and selection filter.
Comparing the microcanonical distribution of $Z_{1}$ for three bombarding energies, we observe no difference in the $E^{*}$ range [2,7] A.MeV where both accuracy of the calorimetry
procedure to deduce excitation energy, detection performance and selection of events of interest are almost under control. This excitation energy range is chosen for the next
step, to obtain information related to density of states of nuclei using fit procedure between canonical and experimental renormalized correlation.
For this purpose, in ref.~\cite{Renorm}, the author proposes an analytical expression for the canonical correlation using the double saddle point approximation.
Considering two phases, it consists of a Taylor expansion around the centroid associated to each phase. This leads to the convolution of two gaussian
distributions with 11 parameters : 8 mean values and standard deviations (2 phases, 2 dimensions), the correlation factor between $Z_{1}$ and $E^*$ and the population of the two phases.
Then the renormalization is applied and the obtained correlation and comparison procedure can be made with data.
In order to minimize the number of parameters, we decide to calculate the correlation factor from the data and to fix the respective population of the two phases in order
to fullfill the transition point description.
In~\cite{BonnetPRL}, the comparison is made for two bombarding energies of Au+Au reaction at 80 and 100 A.MeV with two selection methods described in refs.~\cite{Pichon,BonnetNPA}.
For the four fit procedures, we obtained 4 sets of parameters and localized $E^*$ and $Z_1$ centroids and proposed a first estimation for the latent heat: 
$\Delta E=E_g-E_l=8.1 (\pm0.4)_{stat.} (+1.2 -0.9)_{syst.}$. Employing, respectively, two different selections (two different bombarding energies) allows  to put systematic
(statistical) error bars on these results. From the general point of view, this exhaustive study confirms the robustness and generic properties of the bimodality signal in 
quasi-projectile fragmentation data. On fig.~\ref{fig_01}, best reproduction of the correlation of one data set is shown.

\begin{figure}[!h]
\begin{center}
\includegraphics[scale=0.50]{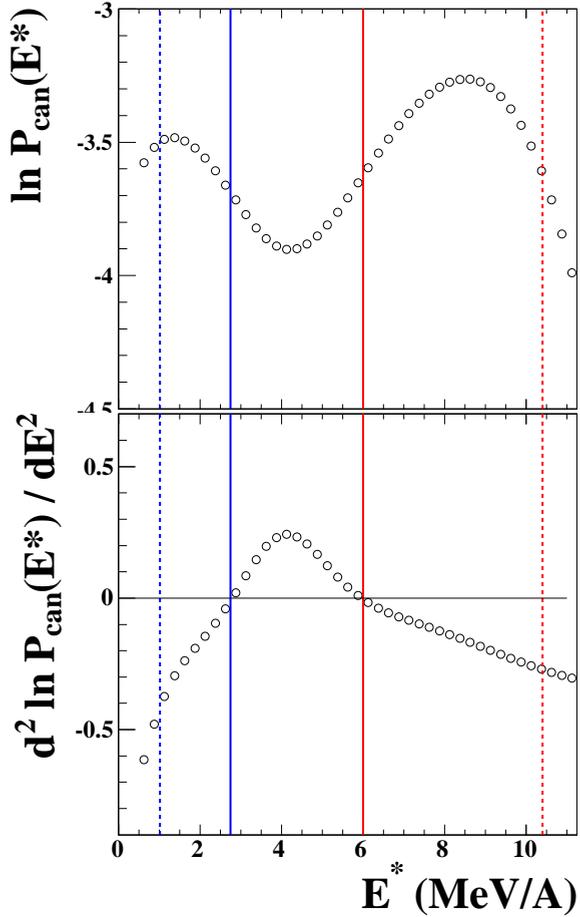}
\caption{\it top figure, logarithme of the projection on the $E^{*}$ axis of the raw canonical 2D correlation $P_{can}(E^{*},Z_{1})$; 
bottom figure, second derivative of the previous one. The dashed lines correspond to the deduce coexistence zone boundaries, the full lines to those of spinodal zone.}
\label{fig_02}
\end{center}
\end{figure}

To go further, a complementary information can be derived from these analyses. From the raw $P_{can}(E^{*},Z_{1}) = W(E^{*},Z_{1})\;e^{-\beta E^{*}}\mathcal{Z}_{\beta}^{-1}$ and look at 
the logarithm of the projection on the $E^{*}$ axis ($d^{2}S/d E^{2} = d^{2} ln\;P_{can}/d E^{2}$). Region where the second derivative
of $ln\;P_{can}(E^{*})$ is positive give us boundaries of the spinodal region. For one of the four set of parameters,
an example of this derivation is shown in fig.~\ref{fig_02}.
In table~\ref{tab_01}, a summary of actual status of experimental boundaries for coexistence and spinodal regions are shown. 
For the spinodal region we have a good coherency between boundaries deduced using bimodality and negative heat capacity signal~\cite{HeatCap}. 

\begin{table}
\begin{center}
\begin{tabular}{|c|c|c|c|}
\hline Boundaries & Signals & Liquid & Gas \\
\hline Spinodal zone\cite{HeatCap} & Negative Heat Capacity & [2.0;2.5] & [5.5;6.5] \\
\hline Spinodal zone & Bimodality & [1.5;3.7] & [6.5;7.0] \\
\hline Coexistence zone\cite{BonnetPRL} & Bimodality & [1;1.7] & [8.5;10.4] \\
\hline
\end{tabular}
\caption{\it Information on the phase diagram region deduced from the study of phase transition signals in quasi-projectiles of Gold.}
\label{tab_01}
\end{center}
\end{table}

\section{Perspectives}

In this work we make an exhaustive comparison between the 2D canonical distribution of two order parameters $E^*$ and $Z_{1}$ and the experimental one.
This direct comparison is possible due to a renormalization procedure and a double saddle point approximation to put at the same level the statistical $E^*$ population.
We extract a value for the latent heat, boundaries of the coexistence zone, and confirm those of the spinodal zone on the excitation energy axis. 
An important information is the evolution of these results with the size of the system. Varying the size/charge of the system will allow to investigate the influence
of Coulomb forces on the stability of nuclei at finite temperature. It will also allow to see the minimum size of the system for which the coexistence of
evaporation and multifragmentation can be observed in terms of bimodality. For these purposes similar analyses on the Xe+Sn collisions at 65, 80 and 100 A.MeV are 
in progress. Another possible experiment could be the exclusive measurement of U+U reaction around the bombarding energy of 100 A.MeV.


%

\end{document}